\documentclass[aps,pra,twocolumn,showpacs,floatfix]{revtex4-2}
\usepackage{graphicx}
\usepackage{amsmath}
\usepackage{amssymb}
\usepackage{color}
\usepackage[separate-uncertainty = true,multi-part-units=single]{siunitx}
\newcommand{\ket}[1]{\left| #1\right\rangle}

\newcommand{\Hetwo}[0]{$\mathrm{He}_2^*$}

\newcommand{\Hefour}[0]{$^4\mathrm{He}$}
\newcommand{\HeII}{He\,II}
\newcommand{\Astate}[0]{$a^3\Sigma_u^+$}

\newcommand{\E}{{\cal E}}
\newcommand{\ILDplus}[0]{I_\updownarrow^\Updownarrow}
\newcommand{\ILDminus}[0]{I_\updownarrow^\Leftrightarrow}
\newcommand{\ICDplus}[0]{I_\circlearrowright^\curvearrowright}
\newcommand{\ICDminus}[0]{I_\circlearrowright^\curvearrowleft}
\newcommand{\LDonethree}[0]{$\text{LD}_{1,3}$}
\newcommand{\CDonethree}[0]{$\text{CD}_{1,3}$}

\begin{document}

\title{Coherent control of molecular rotation in superfluid helium}

\author{Alexander~A.~Milner, Ian~MacPhail-Bartley, Katarina~Preocanin, Shroyon~Dasgupta, Xuanshan~Peng, and Valery~Milner}

\affiliation{Department of  Physics \& Astronomy, The University of British Columbia, Vancouver, Canada}

\date{\today}

\begin{abstract}
We experimentally demonstrate control of molecular rotation in bulk superfluid \Hefour{}. Metastable helium dimers, \Hetwo{}, are rotationally excited by a periodic train of linearly polarized femtosecond laser pulses. We show that the degree of rotational excitation of \Hetwo{} can be enhanced or suppressed by varying the period of the pulse train, whereas the directionality of molecular rotation can be controlled by the relative angle between the polarization vectors of pulses in the train. The experimental results are in agreement with numerical calculations, based on a simple model, in which \Hetwo{} molecules do not interact with the superfluid.
\end{abstract}
\maketitle

This report is a continuation of our recent investigation on the rotational dynamics of metastable helium dimers (\Hetwo{}, also known as helium excimers) in bulk superfluid helium (\HeII{}) \cite{Milner2023b}. Unlike most molecular species, helium excimers can be created and remained dissolved inside bulk \HeII{} under variable temperature and pressure \cite{Surko1968, Dennis1969, Hill1971, Keto1974, Benderskii1999, Mckinsey2003}, serving as native probes of superfluidity in thermodynamic regimes, which are inaccessible to the molecular studies in helium nanodroplets \cite{Toennies2004, Stienkemeier2006}. Ionization of \Hefour{} atoms in the liquid by intense ultrashort laser pulses, followed by a series of recombination processes, results in the production of \Hetwo{} in the lowest metastable triplet state \Astate{} at concentrations of order  \SI[parse-numbers=false]{10^{13}}{cm^{-3}} \cite{Benderskii1999, Milner2023b}. The lifetime of a dimer in the liquid, limited to hundreds of milliseconds due to bimolecular collisions, is sufficiently long to explore its much faster rotational dynamics (rotational periods on the order of a picosecond).

In Ref.~\citenum{Milner2023b}, we have shown that a linearly polarized femtosecond ``kick pulse'' creates a ro-vibrational wave packet in \Hetwo{} - a coherent superposition of states $\ket{v,J}$ with vibrational quantum numbers $v=\{0,1,2\}$, and rotational quantum numbers $J=N,N\pm1$, where $N=\{1,3,5\}$ is the molecular angular momentum excluding the electronic spin. This excitation method is similar to the one used in numerous studies of laser-induced rotational dynamics of molecules both in gas phase \cite{Stapelfeldt2003} and helium nanodroplets \cite{Pentlehner2013, Shepperson2017}. To study these dynamics in bulk \Hetwo{}, we developed a detection technique based on the dichroism in laser-induced fluorescence (LIF). The difference in the absorption of probe pulses of two orthogonal polarizations (detected as the difference between the corresponding LIF signals) indicates the anisotropy of the ensemble-averaged distribution of molecular axes. Hence its dependence on the kick-probe time delay is a measure of the induced rotational coherence \cite{Milner2023b}.

Rather than exciting molecular rotation with a single laser pulse, the ability \textit{to control} rotational wave packets with a sequence of pulses (a ``pulse train'') has been long recognized as a powerful tool in numerous studies of rotational dynamics. It has been used to enhance or suppress molecular alignment in gas samples \cite{Lee2004, Renard2004, Bisgaard2004, Lee2006} and helium nanodroplets \cite{Christiansen2015}, to separate molecular isotopes \cite{Fleischer2006}, isomers \cite{Fleischer2007} and enantiomers \cite{Yachmenev2016, Tutunnikov2018}, to study the chaotic dynamics of quantum rotors \cite{Cryan2009, Zhdanovich2012, Floss2012, Floss2013, Kamalov2015, Bitter2016c}, and to control the directionality of molecular rotation \cite{Fleischer2009, Kitano2009, Zhdanovich2011a, Karras2015}. The desired controllability is achieved due to the quantum interference between the rotational wave packets, excited by individual pulses in the pulse train.

Here, we demonstrate experimentally that rotation of \Hetwo{} molecules dissolved in liquid helium can be controlled by pulse trains in a way similar to the rotational control of gas-phase molecules. Using a sequence of two laser pulses with the same linear polarization (a ``double-kick'' scheme), we show that the rotational excitation is either maintained or fully suppressed, depending on the time separation between the two kicks. Extending the method to longer pulse trains with twisting polarization (a ``chiral pulse train'' method), we demonstrate the induced directionality of the \Hetwo{} rotation. We use a simple model to describe the anticipated signatures of the implemented laser control in the observable dichroism of LIF, and find good agreement between the numerical simulations and the experimental results.

\begin{figure*}[t]
  \includegraphics[width=.9\textwidth]{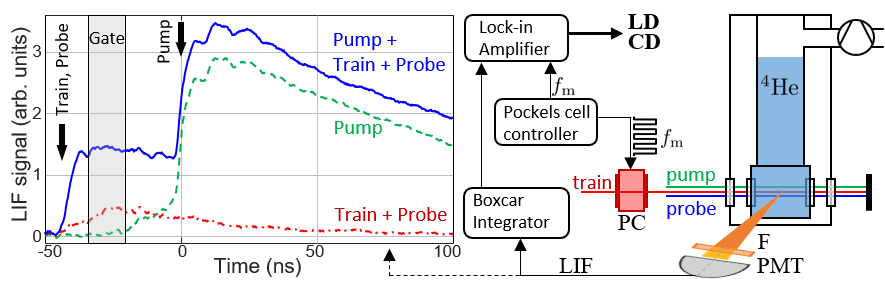}\\
  \caption{Left: examples of the laser-induced fluorescence (LIF) intensity traces, from which the rotation-induced linear or circular dichroism signals (LD and CD, respectively) were deduced. Right: operational diagram of the experimental setup. PC: Pockels cell, PMT: photo-multiplier tube, F: spectral filter.  See text for details.}
  \label{fig-Setup}
\end{figure*}
The experimental setup, schematically shown on the right side of Fig.\ref{fig-Setup}, is similar to that used in our previous work \cite{Milner2023b}. Briefly, the laser beam from a Ti:Sapphire laser system ($\SI{35}{fs}$ pulse length, $\SI{1}{KHz}$ repetition rate, $\SI{793}{nm}$ central wavelength) is split into three parts, hereafter referred to as pump, train and probe. The former is used to produce a sample of \Hetwo{} molecules, whereas the two latter ones are employed to excite and detect the molecular rotation, respectively. The pulses are time delayed with respect to one another, spectrally shaped (as described in detail later in the text), and focused inside a custom-built liquid helium cryostat in a collinear geometry. In the reported series of experiments, we kept the temperature of the liquid at \SI{1.4}{K}, i.e., below the superfluid transition.

The fluorescence is observed at 90 degrees to the excitation direction. It is spectrally filtered around \SI{640}{nm} and focused on a photo-multiplier tube (PMT). The LIF signal is gated in time with a boxcar integrator, as illustrated with a grey rectangle (labeled ``Gate'', drawn to scale) on the left side of Fig.~\ref{fig-Setup}. The plot shows three LIF oscilloscope traces. The dashed green trace represents the fluorescence intensity from the pump pulse alone, arriving at time zero. This fluorescence stems from the radiative decay of the electronically excited excimers, created by the pump ionization of He atoms and the subsequent relaxation processes, leading to the formation of \Hetwo{} \cite{Benderskii1999}. To separate this LIF from the one induced by the probe pulse, the train-probe pair is sent to the sample about \SI{40}{ns} \textit{prior} to the pump (black arrow to the left of the gate box in Fig.~\ref{fig-Setup}). At this time, the excimers (created a millisecond earlier by the previous pump pulse) have long relaxed to their lowest metastable $a$ state, and the fluorescence signal is induced by the train-probe pulse sequence, as shown by the solid blue line in the plot. The red dotted line illustrates the important role of the pump pulses, which increase the steady-state \Hetwo{} concentration and replenish it prior to each rotational excitation/detection window.

As in our earlier work \cite{Milner2023b}, the dichroism signal was measured by modulating the polarization of the pulse train with a Pockels cell (PC), as explained in detail later in the text, while keeping the probe polarization constant. This technique is equivalent to modulating the probe polarization with a constant train, but is less prone to artifacts, related to the static birefringence of optical elements in the setup. The gated LIF signal is fed to the lock-in amplifier, which is tuned to the polarization modulation frequency $f_{m}=\SI{20}{Hz}$, as schematically shown on the right side of Fig.~\ref{fig-Setup}.

\begin{figure*}[t]
  \includegraphics[width=.7\textwidth]{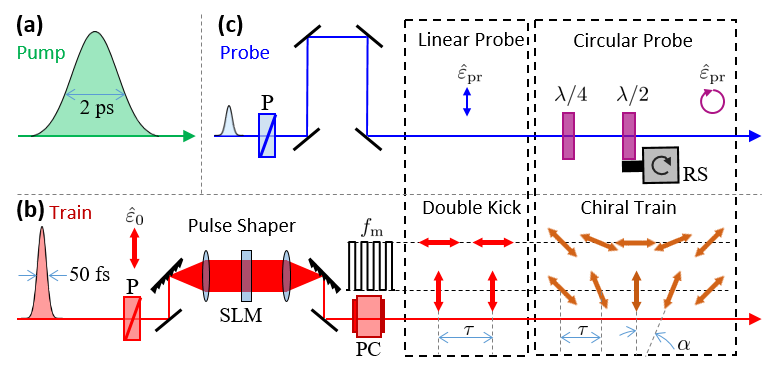}\\
  \caption{Polarization and temporal properties of pump (\textbf{a}), train (\textbf{b}), and probe (\textbf{c}) pulses, used in the reported experiments. See text for details. P: polarizer, SLM: spatial light modulator, $\lambda /2$: half-wave plate, constantly rotating by a motorized rotation stage (RS), $\lambda /4$: quarter-wave plate.}
  \label{fig-Beams}
\end{figure*}

Fig.~\ref{fig-Beams} illustrates the temporal and polarization properties of the three laser beams, sent into the \HeII{} sample. All three beams were focused by a lens with a focal length of \SI{25}{cm} down to $\approx \SI{50}{\mu m}$ beam diameter. Pump pulses are stretched to about \SI{2}{ps} and arrive in the cryostat at the time, designated as time zero in Fig.~\ref{fig-Setup} (black arrow to the right of the gate box). The pump energy of $\approx\SI{100}{\mu J}$ (intensity of \SI{1.7e12}{W/cm^2}) was chosen so as to create the highest possible density of \Hetwo{} molecules, on the order of \SI[parse-numbers=false]{10^{13}}{cm^{-3}}, yet without introducing strong thermal fluctuations and gas pockets, which result in the loss of spatial coherence in the propagation of the laser beams through the liquid \cite{Milner2023b}.

As explained earlier, the excitation pulse train reaches the molecules almost a full millisecond after the pump. The train was produced by means of a pulse shaper, implemented in the standard $4f$ geometry and featuring a liquid crystal-based spatial light modulator (SLM) in its Fourier plane, as depicted in Fig.~\ref{fig-Beams}(\textbf{b}) \cite{Weiner2000}. The shaper converted a linearly polarized transform-limited \SI{50}{fs}-long pulse into either a double-kick train or a chiral pulse train, both with a variable pulse separation $\tau $. In the former case, the two kicks maintain the original polarization $\hat{\varepsilon} _{0}$, whereas in the case of a chiral train, the polarization vector is twisting along the train, i.e., rotates by a variable angle $\alpha $ from pulse to pulse (details on the shaping function to follow).

A Pockels cell was inserted after the pulse shaper, with its optic axis oriented at 45 degrees with respect to $\hat{\varepsilon}_{0}$. The PC is driven by a rectangular waveform, oscillating at the frequency $f_{m}$ between 0 and the half-wave voltage. This periodic modulation either flips the double-kick polarization by 90 degrees, or alternates the handedness of the chiral train, as schematically illustrated in the corresponding dashed rectangles in Fig.~\ref{fig-Beams}(\textbf{b}). The total energy of either pulse sequence was set at $\approx\SI{20}{\mu J}$ (below \SI{6.6e12}{W/cm^2}), which lead to the desired rotational excitation, but did not produce strong fluorescence in the absence of pump light.

Probe pulses of \SI{50}{fs} length and energy around \SI{2}{\mu J} (intensity of \SI{1.3e12}{W/cm^2}) were delayed with respect to the excitation train by a variable amount $\Delta t$, which was scanned between \SI{11}{ps} and \SI{17}{ps}. In the first set of experiments on the double-kick-induced \textit{linear} dichroism (LD), probe polarization was linear and parallel to one of the polarization states of the kicks [left dashed rectangle in Fig.~\ref{fig-Beams}(b)].

The second set of experiments was aimed at demonstrating uni-directional molecular rotation, manifested through the induced \textit{circular} dichroism (CD). Here, we measured the difference in the absorption (again, expressed as the difference in LIF) of a circularly polarized probe for the two chiral trains of opposite handedness. In addition to making the probe polarization circular with a quarter-wave plate, we passed it through a mechanically rotating half-wave plate, as shown inside the right dashed rectangle in Fig.~\ref{fig-Beams}(b). The rotating waveplate served the following purpose. Owing to multiple birefringent optical elements in the setup, such as dielectric mirrors and beam splitters, the polarization of probe pulses is never perfectly circular. The residual ellipticity, even if small, may result in an artificial CD signal due to the linear, rather than circular, anisotropy of the molecular sample. The rotating $\lambda /2$ wave plate scrambles the orientation of the undesired polarization ellipse, averaging the described linear artifacts to zero.

\begin{figure}[b]
  \includegraphics[width=.75\columnwidth]{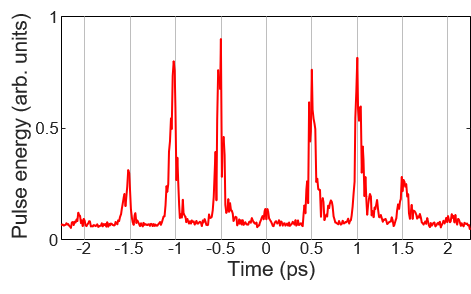}\\
  \caption{Example of an experimental pulse train, used for inducing unidirectional rotation of helium dimers, and measured by means of the standard cross-correlation technique.}
  \label{fig-PulseTrain}
\end{figure}
A femtosecond pulse shaper was used for generating both the double-kick and the chiral train pulse sequences. The former is implemented by applying a simple periodic transfer function, $\cos\left[(\omega -\omega _{0})\tau/2  \right]$, to the spectral field amplitude, centered at frequency $\omega _{0}$ \cite{Weiner2000}. This results in splitting the input pulse into two identical pulses, separated in time by $\tau $.

The details of creating a chiral pulse train are described in our earlier work \cite{Zhdanovich2011a}. Briefly, the two spectral phase masks, $\varphi_{1,2}(\omega )$, of the pulse shaper are modulated according to:
\begin{equation}\label{eq-shaper}
    \varphi_{1,2}(\omega )=A\sin\left[(\omega -\omega _{0})\tau \pm \alpha \right],
\end{equation}
where $A$ is the modulation amplitude, which determines the energy distribution between the pulses in the train, $\tau $ is the train period, and $\alpha $ is the polarization rotation angle between the consecutive pulses in the train. In this work, we chose $A=2.6$, for which the train consists of four pulses of comparable amplitude and a few weaker ones, as shown in Fig.~\ref{fig-PulseTrain}.

As mentioned earlier, we measure the dichroism, either linear (LD) or circular (CD), in the laser-induced fluorescence from the ensemble of rotationally excited \Hetwo{} molecules, which we define as follows:
\begin{equation}\label{eq-LD_CD}
\text{LD}(\Delta t)=\frac{\ILDplus{} - \ILDminus{}}{(\ILDplus{} + \ILDminus{})/2}, \hspace{5mm}
\text{CD}(\Delta t)=\frac{\ICDplus{} - \ICDminus{}}{(\ICDplus{} + \ICDminus{})/2}.
\end{equation}
Here, $\ILDplus{}$ and $\ILDminus{}$ are the fluorescence intensities, recorded with the double-kick polarization (superscript) respectively parallel or perpendicular to the fixed linear probe polarization (subscript). Similarly, $\ICDplus{}$ and $\ICDminus{}$ represent the respective LIF signals for the handedness of the chiral pulse train (superscript) being the same or opposite to the fixed circular probe polarization (subscript).

\begin{figure}[t]
  \includegraphics[width=.99\columnwidth]{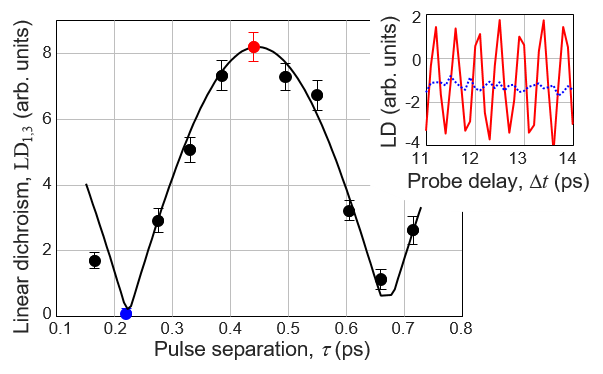}\\
  \caption{Rotation-induced linear dichroism in \Hetwo{} as a function of the pulse-to-pulse separation in the double-kick excitation sequence. Experimental data (filled circles) are compared with the results of numerical simulations (solid line). The dependence of the LD signal on the probe delay is shown in the inset for two values of $\tau $, \SI{220}{fs} (dashed blue) and \SI{440}{fs} (solid red), with the corresponding coloring of the two data points on the main plot. See text for details.}
  \label{fig-LD}
\end{figure}

Owing to the coherent molecular rotation, both the linear and the circular dichroism signals depend on the probe delay $\Delta t$ (see, for example, an oscillatory curve in the inset to Fig.~\ref{fig-LD}). A Fourier transform of this time dependent dichroism shows a strong peak at the frequency of rotational coherence between the two lowest rotational states, $N=1$ and $N=3$, $\nu _{1,3}=\SI{2.27}{THz}$ \cite{Milner2023b}. In what follows, we define the degree of rotational excitation as the amplitude of LD$(\Delta t)$ and CD$(\Delta t)$ at that frequency, denoted as LD$_{1,3}\equiv \text{LD}(\nu _{1,3})$ and CD$_{1,3}\equiv \text{CD}(\nu _{1,3})$, respectively. In the case of the circular dichroism, its sign reflects the directionality of the induced rotation. Below, we present the dependence of LD$_{1,3}$ and CD$_{1,3}$ on the time separation $\tau $ between the pulses in the excitation pulse sequence.

The results of applying a double-kick rotational excitation to helium dimers are shown in Fig.~\ref{fig-LD}. The amplitude of the linear dichroism at the rotational frequency is plotted as a function of the time separation $\tau $ between the two kick pulses. As we varied $\tau $ from \SI{150}{fs} to \SI{715}{fs}, \LDonethree{} exhibited a maximum at \SI{440}{fs}, and two minima at \SI{220}{fs} and \SI{660}{fs}. The former is a result of constructive quantum interference between the two rotational wave packets, created by the two excitation pulses, composed of the $N=1$ and $N=3$ rotational states, and separated by half the rotational period ($1/2\times T_{1,3}=1/\nu _{1,3}$). Similarly, the two minima correspond to the destructive interference at $\tau =1/4\times T_{1,3}$ and $\tau =3/4\times T_{1,3}$.

The inset in Fig.~\ref{fig-LD} shows raw LD signals (Eq.~\ref{eq-LD_CD}) as a function of the probe delay for two values of $\tau $. Strong oscillations at $\tau =\SI{440}{fs}$ (solid red line, corresponding to the red data point in the main plot) indicate strong laser-induced coherent molecular rotation. At $\tau =\SI{220}{fs}$, this rotation is completely turned off, as reflected by the time-independent LD signal (dashed blue line, corresponding to the blue data point in the main plot), which points at the isotropic distribution of molecular axes.

\begin{figure}[t]
  \includegraphics[width=.83\columnwidth]{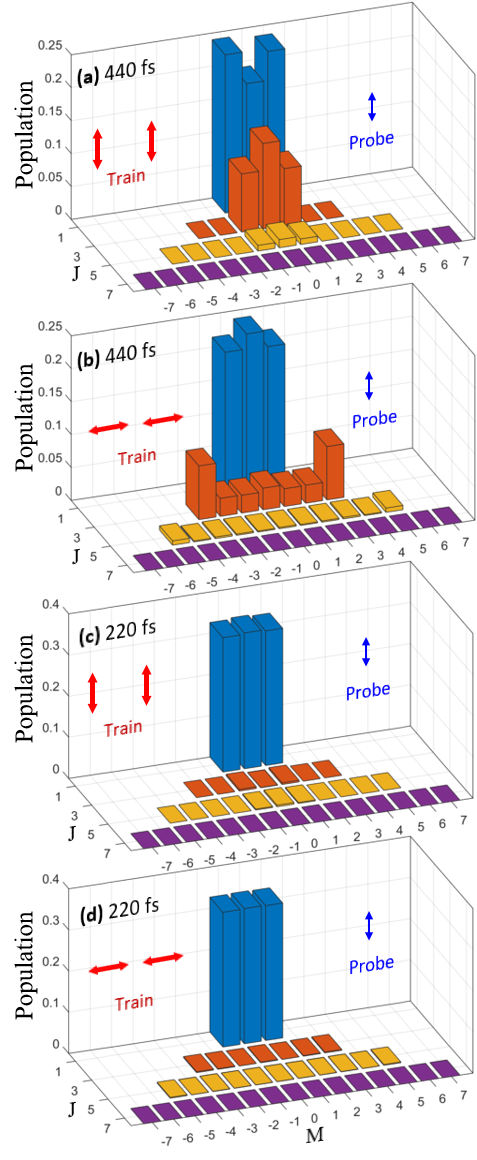}\\
  \caption{Calculated distribution of molecular population among the four lowest rotational levels $J=1,3,5$ and 7, and their corresponding magnetic sublevels $-J\leq M \leq J$. Different panels correspond to different values of pulse separation $\tau $ in a double-kick excitation scheme, and/or different relative orientation of kick and probe pulse polarization vectors, indicated on each plot. See text for details.}
  \label{fig-StatesLinTrain}
\end{figure}
\begin{figure*}[t]
  \includegraphics[width=.75\textwidth]{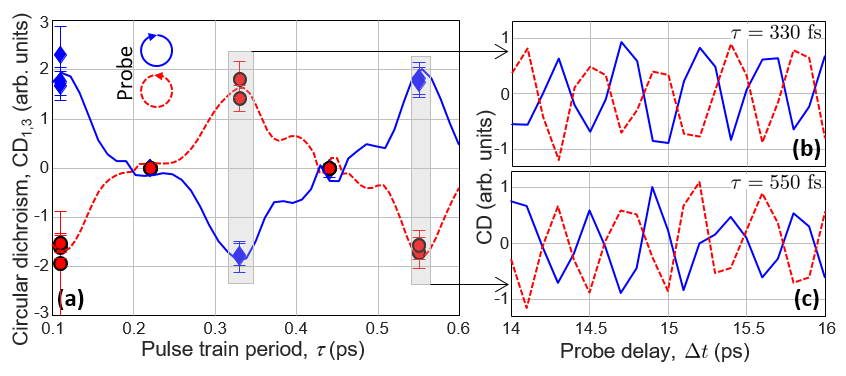}\\
  \caption{(\textbf{a}) Rotation-induced circular dichroism in \Hetwo{} as a function of the period of a chiral pulse train. Experimental data are shown with filled blue diamonds and red circles for the two cases of opposite circular probe polarization (indicated in the upper left corner). The corresponding numerical results are shown with solid blue and red dashed lines, respectively. (\textbf{b,c}) The dependence of the CD signal on the probe delay for $\tau=\SI{330}{fs}$ and $\tau =\SI{550}{fs}$, respectively. Solid blue and dashed red curves correspond to the same circular probe polarizations as in (\textbf{a}). See text for details.}
  \label{fig-CD}
\end{figure*}
Our method of calculating the expected strength of the fluorescence signal, and from that, the value of the rotation-induced dichroism, is the same as in our previous work \cite{Milner2023b}. First, the effect of the excitation pulses, either a double-kick or a chiral pulse train, are described by the induced rotational wave packet:
\begin{equation}\label{eq-wavepacket}
    \psi_{J,M}(t)=\sum_{J,M} c_{J,M} \exp(-iE_J t) \ket{J,M},
\end{equation}
expressed as a coherent superposition of rotational eigenstates $\ket{J,M}$. For a linear rigid rotor, those are the well known spherical harmonics, with $J$ and $M$ being the total angular momentum of the molecule, and its projection on a chosen quantization axis, respectively. The complex state amplitudes $c_{J,M}$ are calculated by numerically solving the Schr\"{o}dinger equation with the interaction Hamiltonian \cite{Floss2012}:
\begin{equation}\label{eq-potential}
V(t)=-\frac{1}{4}\Delta\alpha \cos^2(\theta) \E_{tr}(t),
\end{equation}
where $\Delta \alpha =\SI{35.1}{\AA^3}$ is the difference between the molecular polarizability along and perpendicular to the molecular axis \footnote{J.~Eloranta, private communication. Calculated using the method of Coupled Cluster with Single and Double substitutions, and the basis set from Ref.~\citenum{Eloranta2001}}, $\E_{tr}(t)$ is the field amplitude of the excitation pulse train, and $\theta $ is the angle between the molecular axis and the field polarization. For the initial conditions, all molecules are considered to occupy the lowest rotational level with an isotropic distribution of their axes, as confirmed in the previous report \cite{Milner2023b}.

Fig.~\ref{fig-StatesLinTrain} shows the calculated populations $|c_{J,M}|^{2}$ for a number of interaction scenarios. Note the fixed probe polarization, which also defines the direction of the quantization axis. At the pulse separation of $\tau =\SI{440}{fs}$, the population distributions are quite different between the two orientations of the double-kick polarization, as can be seen by comparing panels (\textbf{a}) and (\textbf{b}). This is the reason for the strong experimentally observed dichroism for this pulse separation (see Fig.\ref{fig-LD}). On the other hand, when $\tau $ is set to \SI{220}{fs}, the second pulse in the double-kick sequence moves all excited population back to the lowest level, regardless of the kicks polarization, as illustrated in panels (\textbf{c}) and (\textbf{d}). This explains the lack of coherent oscillations in the LIF signal (dashed blue line in the inset in Fig.~\ref{fig-LD}), and the correspondingly low value of \LDonethree.

To simulate the experimental signal \LDonethree, we calculated the total population of the excited $d$-state (and hence, the observed fluorescence) under the interaction of the rotational wave packet $\psi_{J,M}(t)$ (determined as explained above) with a weak probe field. In the perturbative regime, this amounts to the real part of the product $c_{1,M}c^*_{3,M}$, summed over all $M$'s with the corresponding two-photon $a\rightarrow d$ transition matrix elements. The result of these calculations is shown by the solid line in Fig.~\ref{fig-LD}, and is in reasonable agreement with our experimental observations. To reproduce the non-zero minimum at \SI{660}{fs}, we added the second vibrational state ($v=1$) to the initial molecular ensemble. Higher vibrational states, owing to the creation mechanism of \Hetwo{} molecules in \HeII{}, were indeed observed in our previous work \cite{Milner2023b}, justifying the interpretation of the incomplete destructive interference as a result of the ro-vibrational dephasing.

\begin{figure*}[t]
  \includegraphics[width=.7\textwidth]{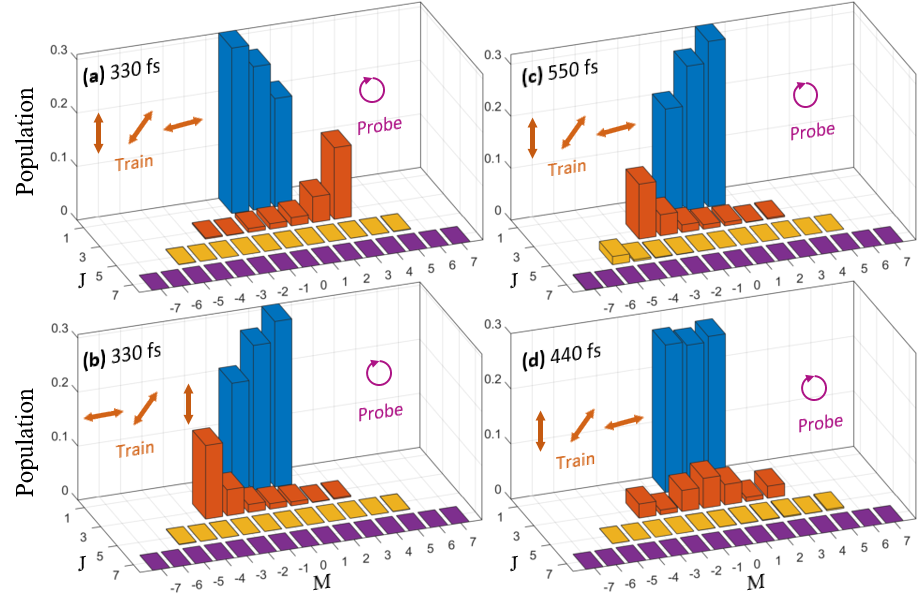}\\
  \caption{Calculated distribution of molecular population among the four lowest rotational levels $J=1,3,5$ and 7, and their corresponding magnetic sublevels $-J\leq M \leq J$. Different panels correspond to different values of the chiral pulse train period $\tau $, and/or different relative handedness of the chiral train and probe pulse polarization. See text for details.}
  \label{fig-StatesChiralTrain}
\end{figure*}
To induce unidirectional rotation of helium excimers, we exposed them to a chiral pulse train. As described earlier in the text, it consists of ~9 equally spaced pulses (six stronger and three weaker ones, see Fig.~\ref{fig-PulseTrain}), whose linear polarization rotates from pulse to pulse by a variable angle $\alpha $. To determine the directionality of rotation, we measured both the amplitude \textit{and sign} of the rotation-induced circular dichroism (\CDonethree{}) with probe pulses of two opposite circular polarizations.

The results are shown in Fig.~\ref{fig-CD}(\textbf{a}). To demonstrate the largest degree of directionality, we used the same chiral train parameters, for which the strongest unidirectional signal has been previously observed in our work in gas samples \cite{Zhdanovich2011a}: the pulse-to-pulse polarization rotation angle of $\alpha =45^{\circ}$ and the pulse train period of either 1/8 or 3/8 of the molecular rotation period. For the former value of $\tau $, the molecular axis follows the field polarization if the molecule rotates in the \textit{same} direction as the field, whereas in the latter case, this happens when the molecule rotates in the \textit{opposite} (to the field) direction. In agreement with this expectation, for a fixed circular probe polarization, e.g., clockwise, \CDonethree{} measured at $\tau =1/8\times T_{1,3} =\SI{110}{fs}$ and $\tau =3/8\times T_{1,3}= \SI{330}{fs}$ exhibited the highest observed amplitude with opposite signs, as indicated by blue diamonds in Fig.~\ref{fig-CD}(\textbf{a}).

Changing the polarization of the probe flips the sign of the circular dichroism [red circles vs blue diamonds in Fig.~\ref{fig-CD}(\textbf{a})]. Raw CD signals as a function of the train-probe delay for two representative values of $\tau $ are plotted on the right side of Fig.~\ref{fig-CD}. The change in the phase of CD$(\Delta t)$ by $\pi $ with either the pulse train period [panel (\textbf{a}) vs panel (\textbf{b})] or the handedness of the probe polarization (solid blue vs dashed red in each panel) confirms the directionality of the induced molecular rotation.

Our numerical calculations of the anticipated circular dichroism in the laser-induced fluorescence from \Hetwo{} support our experimental observations. This is illustrated in Fig.~\ref{fig-StatesChiralTrain}. As in Fig.~\ref{fig-StatesLinTrain}, the vector of probe polarization (here, clockwise circular) was kept constant, and indicated the direction of the quantization axis, with respect to which the magnetic quantum numbers $M$ were determined in the calculations. In contrast to the linearly polarized double-kick, the distribution of the rotational population is not symmetric in $M$, unless the train period is equal to 1/4, 1/2, 3/4, etc. of the rotational period $T_{1,3}$ (see, for example, Fig.~\ref{fig-StatesChiralTrain}(\textbf{d}), where $\tau =1/2\times T_{1,3}=\SI{440}{fs}$, and where both the observed and calculated circular dichroisms are close to zero).

At any other train periods, the population distribution is asymmetric. Moreover, changing the handedness of the chiral train to the opposite, results in flipping the distribution around $M=0$, as can be seen by comparing the distributions in Fig.~\ref{fig-StatesChiralTrain}(\textbf{a}) and (\textbf{b}). The same effect is also caused by changing the train period from $1/8\times T_{1,3}$ to $3/8\times T_{1,3}$, or from $3/8\times T_{1,3}$ to $5/8\times T_{1,3}$ - the latter being apparent in comparing Fig.~\ref{fig-StatesChiralTrain}(\textbf{a}) and (\textbf{c}). Such an asymmetric distribution over the magnetic sublevels is manifest of unidirectional rotation. Its dependence on both the pulse train period and handedness, confirms the interpretation of our experimental results. Indeed, calculating the value of \CDonethree{} (by means of the same numerical procedure, described for the double-kick-induced  \LDonethree{}), we arrive at a good agreement between the experimental observations and numerical results, shown with the solid blue and dashed red lines in Fig.~\ref{fig-CD}(\textbf{a}).

In summary, we experimentally demonstrated coherent control of molecular rotation in superfluid helium by means of femtosecond laser pulse trains. Both the magnitude of the rotational excitation and its directionality were successfully controlled with similar strategies to those used in the gas-phase rotational control. We attribute the high degree of controllability to the weak coupling between \Hetwo{} molecules and the superfluid environment, resulting in correspondingly low decoherence rates, in agreement with our previous findings \cite{Milner2023b}. The ability to control molecular rotation in bulk \HeII{} will be useful for further explorations of superfluidity with molecular probes.

\section*{Acknowledgments}
This research was supported by the Natural Sciences and Engineering Research Council of Canada (NSERC).

%

\end{document}